# Adoption of Blockchain Platform for Security Enhancement in Energy Transaction

[1]Madhuresh Gupta, [1]Soumyakanti Giri, [1]Prabhakar Karthikeyan Shanmugam,
[2,#]Mahajan Sagar Bhaskar [2]Jens Bo Holm-Nielsen, [2]Sanjeevikumar Padmanaban

[1]School of Electrical Engineering, VIT University, Vellore 632014, Tamil Nadu, India.
[2]Center for Bioenergy and Green Engineering, Department of Energy Technology, Aalborg University, Esberg 6700, Denmark.
madhuresh.gupta@outlook.com, giri.soumyakanti@hotmail.com, spk25in@gmail.com,
sagar25.mahajan@gmail.com, san@et.aau.dk, jhn@et.aau.dk

[#]sagar25.mahajan@gmail.com

ABSTRACT:

Renewable energy has become a reality in the present and is being preferred by countries to become a considerable part of the central grid. With the increasing adoption of renewables it will soon become crucial to have a platform which would facilitate secure transaction of energy for consumers as well as producers. This paper discusses and implements a Blockchain based platform which enhances and establishes a secure method to exchange energy. It would also lower the operation costs and accommodate other technologies like the IoT. A basic market mechanism has been developed for peer-to-peer (P2P) transaction of energy where different types of entities can be directly involved. Another concept which is discussed in the paper is the consensus mechanism and whether the model market could hold the security and privacy of the individual users.

## I. Introduction

In today's world where everyone pays high charges for electricity, they can argue the price being charged can be reduced further down with demand-side management for local networks [1]. There are fixed charges [2] which can be reduced considerably by using a blockchain based market due to minimal wastage in case of microgrids and protocols by the supplier side. Keeping the above points in mind, we also need to consider the smart grid scenarios where the individual devices are connected to IoT [3]-[7] which can be managed directly by proper grid protocol and data processing. When all the parameters are considered, it becomes necessary that all the processing is done in a decentralized manner.

Blockchain is a shared ledger that is encrypted and is based on a platform which consists of a network of servers or computers interconnected [8]. Each transaction is validated in the system, also known as mining, which is implemented by doing mathematical calculations to form a block [9]. Based on the type of blockchain, everyone can have access to the same environment, specifically when a public blockchain is considered [10]. Each block is chained or tied to the previous block by embedding the block with information from the previous block [11]. The blockchain is a distributed and decentralized system, which means that it needs to have a way of tracking the official current state of the system. Since the blockchain can include financial transactions and business agreements, it is important that all parties involved are in sync regarding the terms of the agreement.

Since blockchain is decentralized, there is no "higher authority" that can rubber-stamp and finalize the contents of a blockchain block. The architecture is such that no single miner is mining more than half of the total blocks as this would lead to potential manipulations in the data of the blocks added to the chain thereby making it vulnerable to dubious transactions [12]. Thus, to have a balance in the network, no one will be having resources more than half of the resources combined [13]. This is different compared to a traditional transaction where a third party is responsible to verify and authenticate all the transactions as discussed [14]. The comparison of blockchain security with traditional cybersecurity is given in Table I.

TABLE I: Comparison of Blockchain security with traditional cybersecurity

| Blockchain Security | Traditional Cyber Security |
| --- | --- |
| Blockchain works on a decentralized network which is distributed over multiple servers and nodes. | Here, the associated company is completely responsible for the protection and privacy of user data. |
| In Blockchain, the block data is available for every node to verify the transactions and blocks and thereby maintains the integrity of the chain. | Traditional cyber security is highly concentrated on a few servers and systems and the access to these servers are perimeter controlled meaning only a limited number of people have actual access to the data. |
| Having multiple copies of the same chain results in multiple nodes at various locations locked in a chain by hashes making manipulation of data infeasible. | In this form of security, the trusted authorized users in the organization are responsible to maintain the security from external hackers and attackers. |

In this paper, the stability of the grid is considered which can be disrupted due to the integration of multiple renewable energy sources and variability in the production of these sources. The introduction of EV's and battery technology will help in stabilising the grid [15] to a great extent further approving the use of blockchain as a medium. The stability of the grid is key when it comes to the feasibility of a blockchain based grid. Hence, the focus is on creating a platform which is able to execute renewable energy transactions which stays secure from any external attacks.

The paper is organised as follows: Section I focuses on the literature survey, assumptions and the exact outcomes expected from blockchain driven energy market. Further in section II, the implementation of the blockchain platform scenario with two nodes is shown and how it can be achieved in a simplistic and fair manner. In Section III, different cases are elaborated with results and how they might influence the electrical grid. Finally, in Section IV, the conclusion provides a brief outlook on what this research paper does to integrate the P2P energy market [16] with blockchain as a backbone of the platform. The main highlights of the paper are,

- Secure Hash Algorithm (SHA-256) is used as a standard for all types of transactions to protect the chain.
- Everyone is allowed to participate in a common platform to exchange energy.
- Blockchain's direct relation to the electric grid via integration of the P2P market has experimented
- The P2P simulation is discussed in detail.

Blockchain is decentralized and distributed across P2P networks that are continually updated and kept in sync with each other and since they aren't contained in a central location, they don't have a single point of failure and cannot be manipulated from a single computer. It would require massive amounts of computing power to access every instance (or at least a 51% majority) of a certain blockchain and alter them all at the same time.

Thus we have incorporated blockchain for the process of energy exchange among users and keep their data secure from any possible attacks. This also helps to nullify the transaction charges associated with the banks and the fixed charges levied by the energy companies. The structure of a block is defined in a set protocol, wherein the block number, previous block hash [17], transactions of the block, and proof are together hashed to form the hash of the current block.

A major drawback current grid networks is the lack of security in the mannar the transactions are controlled by involvement of mediators and other third parties. This hierarchical organizational trading structure of the grid leads to heavy operating costs with low efficiency of operation [18]. On the other hand, a blockchain-based trading infrastructure offers a decentralized platform that enables the Peer-to-Peer (P2P) trade of energy between consumers and prosumers in a secure manner. The identity privacy and security of transactions is higher in the decentralized platform compared to the traditional system. The P2P energy trade finds purpose in many applications including the Industrial Internet of Things (IIoT) and enhances the possibility of developing micro-grids leading to sustainable energy utilization [19].

II. PROPOSED METHODOLOGY

A sample microgrid with 4 peers was designed to simulate a set of transactions among each other by using blockchain as a common platform to interact and transact energy. The sub-section *A* defines the microgrid used for simulation and sub-section *B* gives a detailed explanation on how the energy is exchanged over blockchain platform.

*A. Model Description*

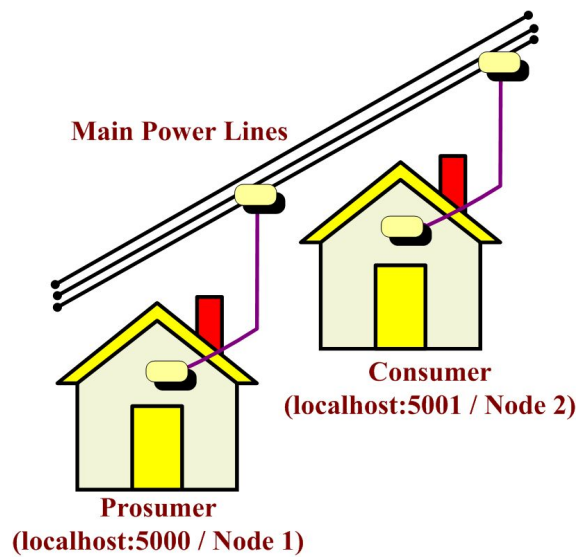

Fig. 1. Sample Microgrid model with 2 peers – 1 consumer and 1 prosumer.

In Fig. 1, two peers have been considered in the microgrid. One of them, a prosumer with solar photovoltaics is considered as an energy producer with the provision to store surplus energy and sell them in the market. Thus, now the consumer has a choice to purchase energy in part from any available seller, the benefit here is a direct exchange of value from one customer to another. The low voltage bus is the general line which will act as the highway for all the exchanges taking place at any given moment. The model consists of 2 nodes. The consumer can directly access the open market space whereas the nodes need to be registered by an internal authentication process. The consensus algorithm depends on the number of people or nodes participating in the network for verifying the transactions. Two virtual nodes are created namely localhost:5000 and localhost:5001 for the sample problem, which has been referenced as one of the consumers and the prosumers. When a transaction occurs, both the nodes keep a history in their respective blocks and in a certain order which is inherent to every chain. In this case, due to the use of two nodes, both the nodes have different ownership in order to prevent a monopoly. The blockchain algorithm for energy transaction is shown in Fig. 2.

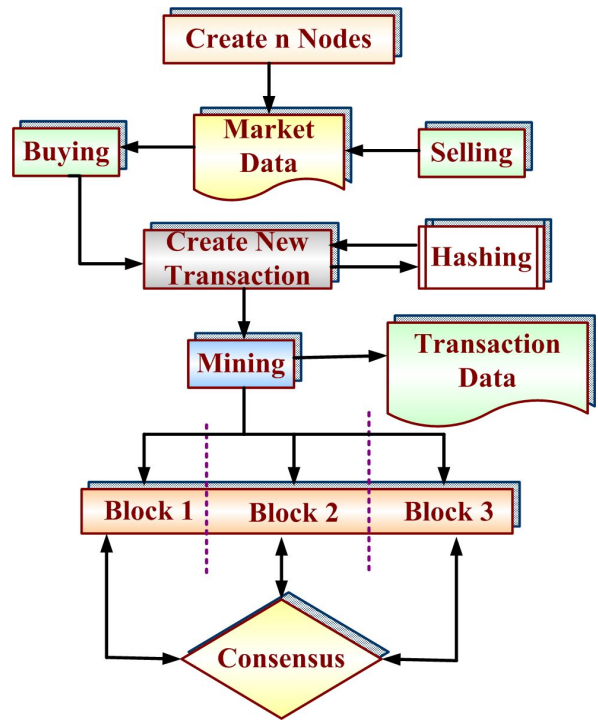

Fig. 2. Blockchain Algorithm for energy transaction.

To simulate a few real-world scenarios, and how blockchain implementation tackles them in contrast to traditional grid energy exchange, the following cases have been discussed:
Case 1: When electrical faults affect the system.
Case 2: When software anomalies happen which could directly affect the grid.

*B. Simulation*
A blockchain class is realised where every block generated will contain the following parameters:
- Multiple transactions
- Hash of the previous block
- Block number
- Timestamp
- Proof (nonce number)

The blockchain concept has been implemented using four sources and two nodes to simulate the real-time transactions which depend on the number of offers posted and the amount quoted by the seller. A user-friendly front-end platform has been developed which simplifies the process of placing energy offers and buying. All the offers available at a particular instant are visible in a tabular form in real-time. The system architecture for P2P transaction is shown in Fig. 3.

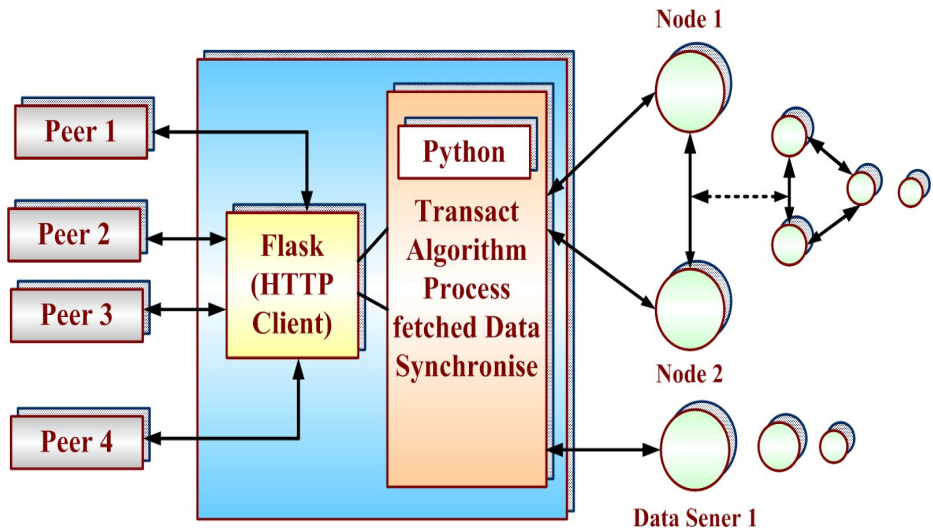

Fig. 3. System architecture for P2P transaction.

First, function *home()* is defined. The *home()* function has a *GET* and *POST* request. The *GET()* request is called whenever the user clicks on the home tab on the navigation bar on the website. The render template method helps in opening the assigned HTML file which in this case is home.html. The home HTML file consists of the main website where all the functionality can be accessed. Next, a function *table()* is defined to initialize the database for storing the transaction exchange data. The *table()* function has a *GET* and *POST* request. The *GET()* request is called whenever the user clicks on the *table tab* on the navigation bar in the website. The *render_template* method helps in opening the assigned HTML file which in this case is *table.html*. The *table html file* consists of all the sellers information required and the price per unit they are offering. The *data.index()* method is used to reiterate the indexes starting from 1. When a user wants to sell energy, they navigate to the *Sell tab* which is rendered by the *sellenergy()* function. When the necessary details are submitted in the seller tab, the user is redirected to the *table tab* in order to see the rest of the pricing. Once the market is set up, the users can *POST()* request from the *home tab*. This returns the required values from another function *buyenergy()* which redirects the user to the page where they can submit the request to buy a certain amount of energy in kWh. After the user has entered the required details in the correct format, the name of the user and the number of units required are called using the *request.form()* method which extracts the necessary data by tracking the form snippet in the HTML file and then matching the name of the form parameter and checking its value entered.

A CSV file named "*energydemand*" is created which stores the current market offerings. This list is read using the *read_csv()* function from the pandas module. Then, check if the required energy by the user is in limits of what the market has to offer. If true, then break the energy required into smaller parts according to what the sellers have to offer. If the energy required is more or equal to the minimum viable amount given by a certain seller then the transaction is recorded between the concerned seller and buyer and the seller is removed from the market. The remaining units, if any, can be acquired from other sellers for which the algorithm needs to recalculate the minimum viable price per unit. When the required units are less than what anyone has to offer, then the seller with the minimum viable amount is contacted and the required units are subtracted from the total units the buyer has to offer. The *iloc* and *loc* methods are used to locate the necessary matrix rows and cells as and when required. After the necessary matchings, all of the transactions in the list are forwarded to class method *new_transactions()* to be added to the next block when mined. The *new_transaction()* method is created in order to account for the sender, the recipient and the amount interchanged between them. These parameters are stored in a dictionary. This dictionary gets appended as a single block to a list each time a new transaction happens in the energy market. The list content increases when the transactions are mined and forged into a new block in the blockchain. After the block is added, the method returns the block number to the calling method. The block created is then passed as an object to the hash function which creates a 256-bit binary number in a 64-bit hexadecimal format. This is done by using the hashlib library. The block is in an object format which has to be converted into a *utf-8 compliant string*. The proof of work algorithm makes use of the hash of the last block, the proof of the last block and the proof which iterates to find a hash with leading four zeros. The *proof_of_work()* method acquires the proof of last block and the hash by calling the hash() property decorator. The two attributes along with the proof are passed to the *valid_proof()* method to check whether the hash created by the combination of these parameters is in terms with what is required. The guess parameter is created which stores a single long string of all the three parameters and then *SHA256* hashing is used in hexadecimal format to find the necessary hash when the proof attribute is iterated by 1 each time the *valid_proof()* method fails to return True. The *mine()* function is a GET request which when called returns the current block information in a *json* format. Then, the last block is called and find the proof of the last block by calling the *proof_of_work()* method inside the blockchain class as discussed previously. After the proof is calculated, a transaction takes place with default parameters to take a certain cut of the transaction because of the mining process. Then call the *new_block()* method to create this block along with the calculated proof. A dictionary response is created which contains the 5 class parameters that are assumed in the beginning. The response is converted to *json* format and returned to make it easy to store in SQL libraries.

### III. RESULTS AND DISCUSSION

A 24-hour average load curve of residential and commercial loads is shown in Fig. 4. This profile gives us an understanding of when the prices are high in the market and when the prices are low.

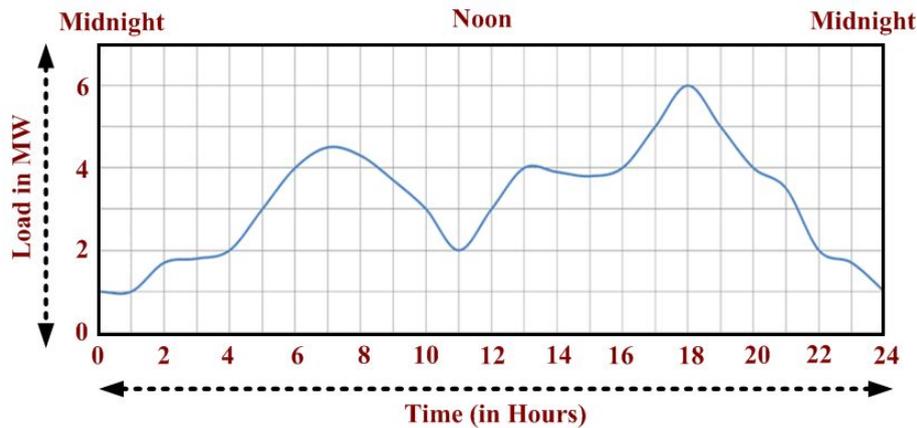

Fig. 4. A 24-hour average load curve of residential and commercial loads [14].

The overall price of electrical energy provided depends on parameters as discussed in [18]. Taking into account these parameters, the prospectus of having a set domain seems a valid point as discussed in the set of equations mentioned.

*X= {ppu, units}*
*0<=ppu< price set by the concerned authority based on the region*(1)
*0<units<= maximum demand of an area*

The change of the boundary values depends on the market competition, consumption pattern, and also according to the government compliance of the respective area, be it anywhere in the world. This domain in the algorithm discussed assures that no one can put arbitrary values or elude with customers who have less knowledge of the market but want to participate.

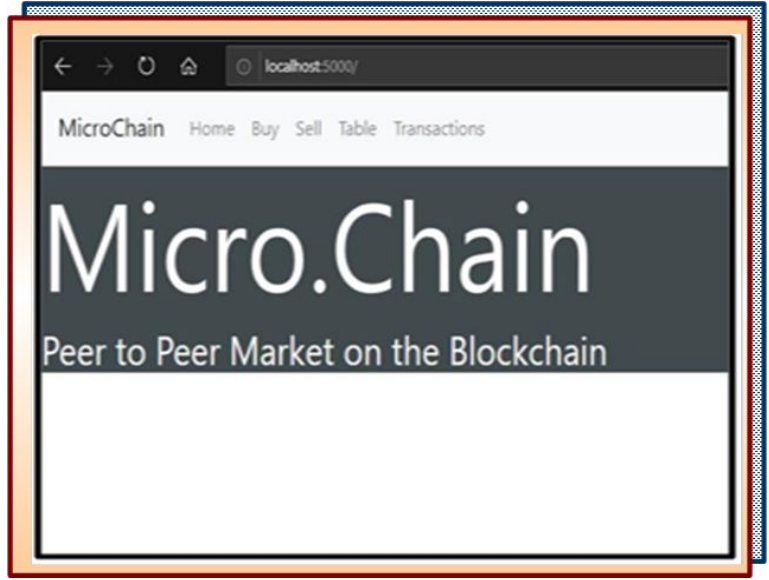

Fig. 5. Frontend of the homepage for Peer-to-Peer market.

In Fig. 5, the frontend of the homepage for Peer-to-Peer market is shown which has multiple links that serve the function of buying, selling, viewing the current listings, and the transactions approved through mining with the completion of power transfer. This interface has been used to simulate the test cases as discussed in the following cases.

*A. Case 1: When electrical faults affect the system*

In this case, electrical faults can cause loss of system memory and failure of live transactions in the blockchain network. This is due to the participation of prosumers and small entities in the market helping in the mining process, if eligible. Their systems cannot be expected to be as robust as the servers in the big industries. Using the blockchain platform, a list of transactions is created which were mined by both the localhosts. But some of the transactions were deliberately left out in the localhost: 5001 to simulate a power failure at one node. The localhost:5000 chain of transaction in json format is given in Table II. The localhost:5001 chain of transaction in json format - data is incomplete is given in Table III. In Table IV, Faulty node- localhost:5001was resolved and replaced with the updated set of transactions. On executing the consensus algorithm, the localhost:5001 with incomplete data block gets compared and replaced by localhost:5000 which has complete database thereby synchronising the transactions which were missing after power failure and electrical fault.

TABLE II. LOCALHOST:5000 CHAIN OF TRANSACTION IN JSON FORMAT

{"chain":
[{"index":1,"previous hash":"1","proof":100,"timestamp":1556980897.851643,"transactions":[]},#
{"index":2,"previoushash":"93f7fce458227e6889abb66fae59809ef8a1b959dcf124da98d2f3129723fc8a
","proof":3859,"timestamp":1556982715.0269084,"transactions":
[{"amount":8.0,"recipient":"Kristian Stromberg","sender": "Tanisha Tichi"},
{"amount":1,"recipient":"68fb958440d949a784d43f59ab4b69f3", "sender":"0"}]},
{"index":3,"previous_hash":"f9b6809af6bffe1bbd99643f8b3b4a269e4b20ba8d89fb7f363efe5a6333e507

","proof":8486,"timestamp":1556983172.057091, "transactions":
[{"amount":4.0,"recipient": "Kristian Stromberg", "sender": "Ellis Acost"},
{"amount":5.0,"recipient": "Apryl Goulet", "sender":"Ellis Acost"},
 {"amount":1,"recipient":"68fb958440d949a784d43f59ab4b69f3", "sender":"0"}]},
 "index":4,"previoushash":"8047bc@ab2748566293a55f29b94f42a366855b8dfd6ce97f60d517d679f3f5
,"proof":10419,"timestamp":1556983962.120725,"transactions":
[{"amount":5.0,"recipient":"Apryl Goulet", "sender":"Chisel Acincio"},
 {"amount":1,"recipient":"68fb958440d949a784d43f59ab4b69f3", "sender":"0"}]}],"length":4}

TABLE III. LOCALHOST:5001 CHAIN OF TRANSACTION IN JSON FORMAT - DATA IS INCOMPLETE

{"chain":
[{"index":1,"previous_hash": "1", "proof":100,"timestamp":1556981581.5241313,"transactions":
[]},
{"index":2,"previous hash":"6f08dcb23c4475d28b1c8bda2669c045952a0ec36e6a6fbf419193e0a9e62b6a
","proof":102785,"timestamp":1556982856.725807,"transactions":
[{"amount":8.0,"recipient":"Kristian Stromberg","sender": "Tanisha Tichi"},
 {"amount":1,"recipient":"f54793eb52714db49632a7599aa87f28", "sender":"0"}]},
{"index":3,"previous hash":"0ab8cec5a21f9c76deac8882f54a675e2ce178ced4b42859ac42118ddd46376a
","proof":21331,"timestamp":1556983221.1510096,"transactions":
[{"amount":4.0, "recipient":"Kristian Stromberg", "sender": "Ellis Acost"},
 {"amount":5.0,"recipient":"Apryl Goulet", "sender": "Ellis Acost"},
{"amount":1,"recipient": "f54793eb52714db49632a7599aa87f28", "sender": ""}]}],"length":3}

TABLE IV. FAULTY NODE- LOCALHOST:5001 WAS RESOLVED AND REPLACED WITH THE UPDATED SET OF TRANSACTIONS

"message": "our chain was replaced",
"new_chain": [
        {"index": 1,
        "previous_hash": "1",
        "proof": 100,
        "timestamp": 1556980897.851643,
        "transactions": []
        },
        {index": 2,
        "previous_hash":"93f7fce458227e6889abb06fae59809ef8a1b959dcf124da98d2f3129723fc8a",
        "proof": 3859,
        "timestamp": 1556982715.0269084,
        "transactions": [
        },

*B. Case 2: When software anomalies happen which could directly affect the grid*

This is a scenario which could happen when users are not doing an integrity check of their software regularly, or due to tampering of transactions for personal benefit. If not resolved quickly, multiple anomalies can break the microgrid protocols by changing energy patterns in the network. So in order to simulate such a situation, the same list of transactions are used but in this case, instead of technical shortcomings, then change the underlying data of localhost:5001 with the length of chain same as

localhost:5000. A set of transactions is also saved in the local server which can be used as a comparison along with localhost:5000 to verify which chain is true and replace the false with the actual one. As seen in Fig. 6, both the server data and the chain in localhost:5000 have the same data.

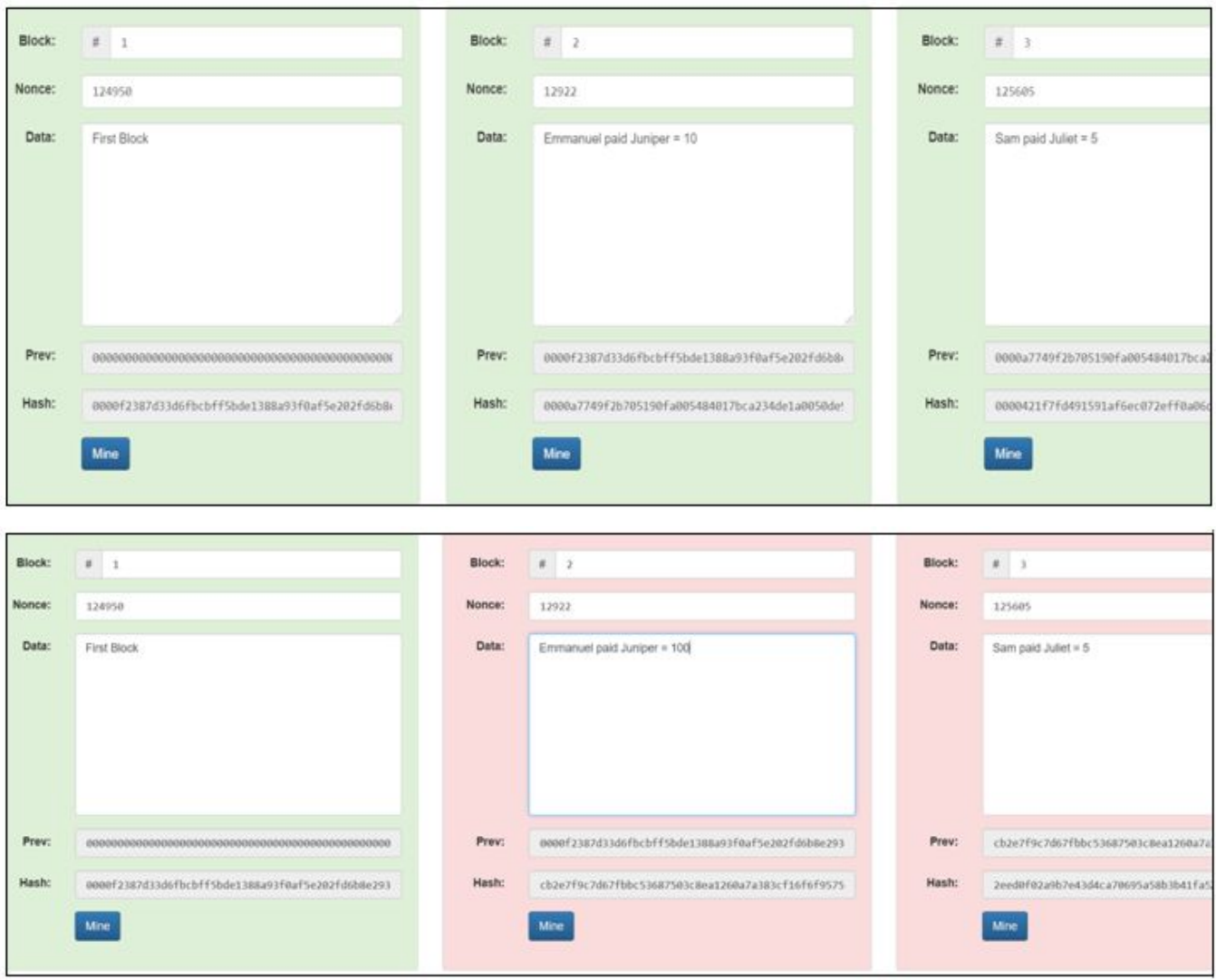

Fig. 6. The second chain breaks due to alteration in block data (indicated by pink blocks)

In Fig. 6, we are showing two similar blockchains (representing different nodes of the network) so to represent someone trying to manipulate the second chain by adding a fraudulent transaction, we can observe how the chain breaks right from the block where the data was manipulated. Considering the two cases approached above, and the algorithm which aids the simulation, it can be seen how the effect of one of the systems fails does not have instant feedback on the system parameters and yet, when the system works perfectly, the transactions are comparable to the security and speed where intermediaries are involved. This delay in change helps the system to revert back to a healthy status without having any effect on the software, specifically the users and their wallets, referring to the transactional anomalies. Also, the grid shows anomaly in power transfers which could be a bigger problem as it could try to overload users with limited to no battery storage and create unnecessary faults in the grid which need to be manually repaired causing loss of money, comfort and time for the locals. In Table V, consensus algorithm replaced the broken (altered) chain with the correct chain automatically.

TABLE V. CONSENSUS ALGORITHM REPLACED THE BROKEN (ALTERED) CHAIN WITH THE CORRECT CHAIN AUTOMATICALLY

"message": "our chain was replaced",
"new_chain": [
{"index": 1, "previous_hash": "1",
  "proof": 100,
  "timestamp": 15588088924.316836,
  "transactions": []},
{ index": 2,
  "previous_hash": "cca9c11299562ec87576eb1a10e8197543228ff08481666ed2a760fa80f269c4",
  "proof": 48714,
  "timestamp": 1558809772.212345,
"transactions": [
  {"amount": 10,
"recipient": "Apryl Goulet",
"sender": "Juniper Charlotte"

Electrical grid integrity can be maintained locally, with the local transactions driving the algorithm for how smart grids behave. This would be much easier than extracting from a central server each time a small distance and small power transactions needs to take place which would be more common. The long-distance, large energy transactions would require a dedicated area-specific server to process data for which is considered in the form of a CSV file which keeps updating. These types of servers would be most useful in aiding the P2P transactions indirectly by making the grid calculations themselves without putting pressure on the nodes that are already busy in verifying the transactions and transferring power depending on the contract obligations. The results obtained prove that the blockchain model can bring out the meaning of prosumers in true sense wherein, the large company is at the same level in terms of price offerings. The product can be deployed at scale with little physical infrastructure, and henceforth, more and more servers can be simultaneously added for better reliability. There is no distinction between the large power providers and the small power provider, be it 1kWh seller or 1 MWh seller. This also means that the data which was kept by the large companies no longer belongs to them as they are equal. It belongs to everybody participating in the market and here blockchain creates an easy way to keep track of everything happening around the market open for every participant without any bias.

## IV. CONCLUSIONS

This paper deals with the proposal of a P2P system using blockchain that enables it to maintain stability and uniformity in the grid. A very lightweight architecture was also created as to how the application can be handled effectively with few resources to be used upon the node client and user side. In particular, the paper focuses on how blockchain directly relates to the stability of the microgrid architecture. The sample demonstration has shown how the implementation proves to be robust and reliable in storing and performing transactions. This maintains the integrity of the local grid even in adverse conditions. The paper briefly discusses how the pricing will be affected based on the location and the consumption patterns. Overall, this paper gives a detailed insight into how a P2P transaction system is created on the principle of blockchain and how it benefits the maintenance and control of the next-gen grid in the long run.

## REFERENCES

[1]. Li, X. H., & Hong, S. H. (2014). User-expected price-based demand response algorithm for a home-to-grid system. Energy, 64, 437-449.

[2]. Tarbi, L. (2019). The Simple Way to Read Your Electric Bill | EnergySage. [online] Solar News. Available at: https://news.energysage.com/whats-the-right-way-to-read-your-electric-bill/ [Accessed 11 Mar. 2019].
[3]. G. Ghatikar, "INTERNET OF THINGS AND SMART GRID STANDARDIZATION," in *Internet of Things and Data Analytics Handbook*, H. Geng, Ed. Hoboken, NJ, USA: John Wiley & Sons, Inc., 2016, pp. 495–512.
[4]. A. Ahmed Abdulkadir and F. Al-Turjman, "Smart-grid and solar energy harvesting in the IoT era: An overview," *Concurrency and Computation: Practice and Experience*, Wiley, p. e4896, Aug. 2018.
[5]. D. Minoli and B. Occhiogrosso, "The Emerging 'Energy Internet of Things,'" in *Internet of Things A to Z*, Q. Hassan, Ed. Hoboken, NJ, USA: John Wiley & Sons, Inc., 2018, pp. 385–424.
[6]. N. S. Srivatchan and P. Rangarajan, "A novel low-cost smart energy meter based on IoT for developing countries' micro grids," *Concurrency and Computation: Practice and Experience*, Wiley, p. e5042, Oct. 2018.
[7]. Huh, S., Cho, S., & Kim, S. (2017, February). Managing IoT devices using blockchain platform. In 2017 19th international conference on advanced communication technology (ICACT) (pp. 464-467). IEEE.
[8]. Crosby, M., Pattanayak, P., Verma, S., & Kalyanaraman, V. (2016). Blockchain technology: Beyond bitcoin. Applied Innovation, 2(6-10), 71.
[9]. Vukolić, M. (2015, October). The quest for scalable blockchain fabric: Proof-of-work vs. BFT replication. In International workshop on open problems in network security (pp. 112-125). Springer, Cham.
[10]. Zheng, Z., Xie, S., Dai, H., Chen, X., & Wang, H. (2017, June). An overview of blockchain technology: Architecture, consensus, and future trends. In 2017 IEEE International Congress on Big Data (BigData Congress) (pp. 557-564). IEEE.
[11]. Gabison, G. (2016). Policy considerations for the blockchain technology public and private applications. SMU Sci. & Tech. L. Rev., 19, 327.
[12]. Chen, L., Xu, L., Gao, Z., Lu, Y., & Shi, W. (2018, June). Protecting Early Stage Proof-of-Work Based Public Blockchain. In 2018 48th Annual IEEE/IFIP International Conference on Dependable Systems and Networks Workshops (DSN-W) (pp. 122-127). IEEE.
[13]. Puthal, D., & Mohanty, S. P. (2019). Proof of Authentication: IoT-Friendly Blockchains. IEEE Potentials, 38(1), 26-29.
[14]. Puthal, D., Malik, N., Mohanty, S. P., Kougianos, E., & Das, G. (2018). Everything you wanted to know about the blockchain: Its promise, components, processes, and problems. IEEE Consumer Electronics Magazine, 7(4), 6-14.
[15]. Gupta, M., &Giri, S. "Impact of Vehicle-to-Grid on Voltage Stability - Indian Scenario." 2018 National Power Engineering Conference (NPEC), 2018, doi:10.1109/npec.2018.8476749.
[16]. Park, C., & Yong, T. (2017). Comparative review and discussion on P2P electricity trading. Energy Procedia, 128, 3-9.
[17]. Dworkin, M. J. (2015). SHA-3 standard: Permutation-based hash and extendable-output functions (No. Federal Inf. Process. Stds.(NIST FIPS)-202).
[18]. Abdella, J., & Shuaib, K. (2018). Peer to peer distributed energy trading in smart grids: A survey. Energies, 11(6), 1560.
[19]. Mengelkamp, E., Notheisen, B., Beer, C., Dauer, D., & Weinhardt, C. (2018). A blockchain-based smart grid: towards sustainable local energy markets. Computer Science-Research and Development, 33(1-2), 207-214.
[20]. Jardini, J. A., Tahan, C. M., Gouvea, M. R., Ahn, S. U., &Figueiredo, F. M. (2000). Daily load profiles for residential, commercial and industrial low voltage consumers. IEEE Transactions on power delivery, 15(1), 375-380.